\documentclass[useAMS,usenatbib]{mn2e}

\usepackage{epsfig}
\usepackage{amsmath}
\usepackage{amssymb}
\usepackage{mathrsfs}
\usepackage{graphicx}

\newcommand{\apj}{ApJ} 
\newcommand{\apjl}{ApJ} 
\newcommand{\apjs}{ApJ} 
\newcommand{\aj}{AJ} 
\newcommand{\aap}{A\&A} 
\newcommand{\araa}{ARAA} 
\newcommand{\mnras}{MNRAS} 
\newcommand{\nat}{Nature} 

\newcommand{\mach}{\mathcal{M}}
\newcommand{\rhothreev}{\rho^{1/3}v}
\newcommand{\sigs}{\sigma_s}
\newcommand{\kinj}{k_\mathrm{inj}}
\newcommand{\ks}{k_\mathrm{s}}
\newcommand{\deriv}{\mathrm{d}}

\newcommand{\vect}[1]{{\textbf{\textit{#1}}}}
\newcommand{\cs}{c_\mathrm{s}}

\title[Supersonic turbulence]{On the universality of supersonic turbulence}

\author[Federrath]{Christoph~Federrath$^{1}$
\thanks{E-mail: christoph.federrath@monash.edu}\\
$^{1}$Monash Centre for Astrophysics, School of Mathematical Sciences, Monash University, VIC 3800, Australia}

\begin{document}



\maketitle


\begin{abstract}
Compressible turbulence shapes the structure of the interstellar medium of our Galaxy and likely plays an important role also during structure formation in the early Universe. The density probability distribution function (PDF) and the power spectrum of such compressible, supersonic turbulence are the key ingredients for theories of star formation. However, both the PDF and the spectrum are still a matter of debate, because theoretical predictions are limited and simulations of supersonic turbulence require enormous resolutions to capture the inertial-range scaling.
To advance our limited knowledge of compressible turbulence, we here present and analyse the world's largest simulations of supersonic turbulence. We compare hydrodynamic models with numerical resolutions of $256^3$--$4096^3$ mesh points and with two distinct driving mechanisms, solenoidal (divergence-free) driving and compressive (curl-free) driving. We find convergence of the density PDF, with compressive driving exhibiting a much wider and more intermittent density distribution than solenoidal driving by fitting to a recent theoretical model for intermittent density PDFs. Analysing the power spectrum of the turbulence, we find a pure velocity scaling close to Burgers turbulence with $P(v)\propto k^{-2}$ for both driving modes in our hydrodynamical simulations with Mach number $\mathcal{M}=17$. The spectrum of the density-weighted velocity $\rho^{1/3}v$, however, does not provide the previously suggested universal scaling for supersonic turbulence. We find that the power spectrum $P(\rhothreev)$ scales with wavenumber as $k^{-1.74}$ for solenoidal driving, close to incompressible Kolmogorov turbulence ($k^{-5/3}$), but is significantly steeper with $k^{-2.10}$ for compressive driving. We show that this is consistent with a recent theoretical model for compressible turbulence that predicts $P(\rhothreev)\propto k^{-19/9}$ in the presence of a strong $\nabla\cdot\vect{v}$ component as is produced by compressive driving and remains remarkably constant throughout the supersonic turbulent cascade.
\end{abstract}


\section{Introduction}

The aim of this study is to pin down the properties and statistics of supersonic, compressible turbulence. This kind of turbulence is relevant for the highly compressible interstellar medium \citep{MacLowKlessen2004,ElmegreenScalo2004,McKeeOstriker2007}, because it controls the rate of star formation triggered by gas compression in shocks \citep{KrumholzMcKee2005,PadoanNordlund2011,HennebelleChabrier2011,FederrathKlessen2012}, affects the star formation efficiency \citep{Elmegreen2008,FederrathKlessen2013,KainulainenFederrathHenning2013}, and determines the mass distribution of stars when they are born \citep{PadoanNordlund2002,HennebelleChabrier2008,HennebelleChabrier2013,Hopkins2013IMF}. Supersonic turbulence also has an important effect on the gravitational instability of galactic discs \citep{RomeoBurkertAgertz2010,HoffmannRomeo2012}. Even the early Universe was likely dominated by supersonic turbulence when the first cosmic haloes started to contract to form the first galaxies \citep{AbelBryanNorman2002,GreifEtAl2008,WiseTurkAbel2008,SchleicherEtAl2010}. Analytic models of star formation are based upon the probability distribution function (PDF) of the gas density and the scaling of the velocity spectrum in supersonic turbulence. It is thus crucial to determine the PDF and the scaling with high precision and to test whether these are universal in any kind of supersonically turbulent flow or whether they depend on the driving of the turbulence.

It is important to study the influence of the driving mode, because interstellar turbulence is likely driven by a combination of different stirring mechanisms, all leading to potentially different excitation states and mode mixtures. Driving mechanisms for interstellar turbulence include supernova explosions and expanding, ionizing shells from previous cycles of star formation \citep{McKee1989,KrumholzMatznerMcKee2006,BalsaraEtAl2004,BreitschwerdtEtAl2009,PetersEtAl2011,GoldbaumEtAl2011,LeeMurrayRahman2012}, gravitational collapse and accretion of material \citep{VazquezCantoLizano1998,KlessenHennebelle2010,ElmegreenBurkert2010,VazquezSemadeniEtAl2010,FederrathSurSchleicherBanerjeeKlessen2011,RobertsonGoldreich2012,ChoiShlosmanBegelman2013}, and galactic spiral-arm compression of H\textsc{I} clouds \citep{DobbsBonnell2008,DobbsEtAl2008}, as well as magneto-rotational instability \citep{PiontekOstriker2007,TamburroEtAl2009}. Wind, jets and outflows from young stellar objects have also been suggested to drive turbulence on smaller scales \citep{NormanSilk1980,BanerjeeKlessenFendt2007,NakamuraLi2008,CunninghamEtAl2009,CarrollFrankBlackman2010,WangEtAl2010,MoraghanKimYoon2013}. Turbulence in high-redshift galaxies is probably generated during the collapse of primordial haloes and later by the feedback from the first stars \citep{GreifEtAl2008,GreenEtAl2010,LatifEtAl2013}.

Many of the aforementioned driving mechanisms for interstellar turbulence directly compress the gas (which we call `compressive driving'), while others primarily excite vortices (called `solenoidal driving'). Mathematically, we distinguish those two extreme cases by defining a vector field $\vect{F}_\mathrm{stir}$ that drives the turbulence \citep{FederrathKlessenSchmidt2008,FederrathDuvalKlessenSchmidtMacLow2010}:
\begin{itemize}
\item solenoidal driving ($\nabla\cdot\vect{F}_\mathrm{stir}=0$), and
\item compressive driving ($\nabla\times\vect{F}_\mathrm{stir}=0$).
\end{itemize}
In reality, we expect a mixture of both, some mechanisms will be closer to our idealized picture of solenoidal driving, while others might be closer to compressive driving.

Unlike the extensively studied case of incompressible turbulence led by the pioneering theoretical work of \citet[][hereafter K41]{Kolmogorov1941c}, studies of highly supersonic turbulent flows only recently started to shed light on the basic statistics of supersonic turbulent flows. Because of its complexity and three-dimensional nature, the properties of supersonic, compressible turbulence are primarily investigated through numerical simulations \citep[e.g.,][]{PorterPouquetWoodward1992,KritsukEtAl2007,SchmidtEtAl2009,FederrathDuvalKlessenSchmidtMacLow2010}. Early studies \citep{PorterPouquetWoodward1994} indicated that compressible turbulence might exhibit a turbulent velocity spectrum $P(v)$ very similar to the phenomenological theory of incompressible turbulence by Kolmogorov with $P(v)\propto k^{-5/3}$ \citep[K41,][]{Frisch1995}. Here, $v$ is the turbulent gas velocity and $k=2\pi/\ell$ is the wavenumber (or inverse length-scale $\ell$) of a turbulent fluctuation (sometimes called `eddy'). The resolution of these early simulations, however, did not yield a significant inertial range (the scaling range over which a power law in wavenumber space can be measured which is well separated from both the driving and the viscous scales), and the turbulence was only mildly compressible \mbox{(Mach~number~$\lesssim1$)}.

With the advent of supercomputers combining thousands of compute cores in one large-scale parallel application, it is only recently that the spectral scaling of supersonic turbulence could be measured with improved precision \citep{KritsukEtAl2007,FederrathDuvalKlessenSchmidtMacLow2010}, indicating $P(v)\propto k^{-2}$, which is much steeper than the Kolmogorov spectrum and closer to Burgers turbulence \citep{Burgers1948}. Burgers turbulence consists of a network of discontinuities (shocks), which can only form in supersonic flows. However, the studies by \citet{KritsukEtAl2007} and \citet{FederrathDuvalKlessenSchmidtMacLow2010} were limited to $1024^3$ grid cells. The highest resolution simulation of supersonic turbulence so far was done by \citet{KritsukEtAl2009} for a moderate Mach number of 6. Although this is clearly in the supersonic regime, typical molecular clouds in the Milky Way have Mach numbers of about \mbox{$5$--$20$} and sizes in the range \mbox{1--$50\,\mathrm{pc}$} \citep[e.g.,][]{RomanDuvalEtAl2010}. They can thus be significantly more compressible. Here we focus on the most compressible type of clouds in the Milky Way and compare simulations with Mach 17 turbulence. \citet{KritsukEtAl2009} only studied solenoidal (divergence-free) driving, while here we study both extremes: solenoidal and compressive (curl-free) driving, in order to test the influence of different driving modes. We find significantly different statistics for these two extreme cases.

First, we briefly summarize our limited theoretical knowledge of supersonic turbulence in Section~\ref{sec:theory}. In Section~\ref{sec:methods}, we then turn to the numerical simulation techniques used to compare to and test these theories. Section~\ref{sec:results} presents our results with details on the vorticity production and spatial structure of supersonic turbulence, the density PDF, and finally the scaling of the power spectrum. We conclude in Section~\ref{sec:conclusions}.

\section{Theory of compressible turbulence} \label{sec:theory}

Studying turbulence requires a sufficient scale separation between energy injection (driving) on large scales and dissipation on small scales. The range in between is known as the inertial range of turbulence with a constant energy flux, where the flow is directly influenced neither by driving nor by dissipation. The existence of an inertial range is well established for incompressible turbulence \citep{Frisch1995}. However, this may not be the case for supersonic turbulence. It is only recently that \citet{Aluie2011,Aluie2013} have rigorously proven the existence of an inertial range for highly compressible turbulence produced by any type of driving mechanism, solenoidal or compressive. The existence of such an inertial range, however, does not exclude the possibility of different scaling properties for solenoidal or compressive driving, which we will test below.

A fundamental idea for the scaling of supersonic turbulence was proposed by \citet{Lighthill1955} and later refined by \citet{Henriksen1991}, \citet{Fleck1996} and \cite{KritsukEtAl2007}. Based on the dimensional analysis by K41 and the assumption of a constant flux of the kinetic energy density, $e_\mathrm{kin}=(1/2)\rho v^2$ in the inertial range, we can write
\begin{equation} \label{eq:flux}
\frac{\deriv e_\mathrm{kin}}{\deriv t} \propto \frac{\rho v^2}{t} \propto \frac{\rho v^3}{\ell} \overset{!}{=} \mathrm{const.}
\end{equation}
The second proportionality implies a time-scale $t=\ell/v$ for energy transfer on scale $\ell$. The last, enforced equality in Equation~(\ref{eq:flux}) is that of a constant energy flux and is the same as that assumed in K41, only that we keep the dependence on density $\rho$, while $\rho=\mathrm{const}$ in the incompressible model by K41. Taking this last equality in Equation~(\ref{eq:flux}), we find that the third-order structure function of the density-weighted velocity, $\vect{v}_\mathrm{mw}\equiv\rho^{1/3}\vect{v} \propto \ell^{1/3}$, scales linearly, $\langle\left|\delta\vect{v}_\mathrm{mw}(\ell)\right|^3\rangle\propto\rho v^3\propto\ell$, for an increment $\delta\vect{v}_\mathrm{mw}(\ell)$ between two points separated by a distance $\ell$. Thus, the original Kolmogorov scaling for the power spectrum
\begin{equation} \label{eq:k07scaling}
P(\rhothreev)\propto \deriv (\rhothreev)^2/\deriv k\propto k^{-5/3}
\end{equation}
would be preserved even for highly compressible turbulence, if the density-weighted velocity $\rhothreev$ were taken instead of the pure velocity $v$. Indeed, numerical simulations with with Mach numbers of \mbox{$\mach\approx5$--$7$} and resolutions of $1024^3$ grid cells, using solenoidal or weakly compressive driving, indicate a $\rhothreev$ scaling consistent with $P(\rhothreev)\propto k^{-5/3}$ \citep{KritsukEtAl2007,FederrathDuvalKlessenSchmidtMacLow2010,PriceFederrath2010,KritsukWagnerNorman2013}, even if a magnetic field is included \citep{KowalLazarian2007,KritsukEtAl2009}. However, the simulation with purely compressive driving by \citet{FederrathDuvalKlessenSchmidtMacLow2010} indicated a significantly steeper scaling with $P(\rhothreev)\propto k^{-2.1}$ in the inertial range.

While this last result might be regarded as a false alarm, because of a limited or insufficient scaling range in simulations with purely compressive driving as argued by \citet{KritsukEtAl2010}, recently, \citet{GaltierBanerjee2011} derived an exact relation for the scaling of compressible isothermal turbulence, which does exactly predict $P(\rhothreev)\propto k^{-19/9}\approx k^{-2.1}$. Their model is also based on $\rhothreev$ and the predicted $P(\rhothreev)\propto k^{-19/9}$ scaling applies for turbulence with a very strong $\nabla\cdot\vect{v}$ component, such as produced by compressive driving. Only around the sonic scale, where the local Mach number drops to unity, would the spectrum approach $P(\rhothreev)\propto k^{-5/3}$.

The central result in \citet{GaltierBanerjee2011} is an exact relation for compressible turbulence (their Equation~11). It has two contributions to the total energy injection or dissipation rate $\varepsilon$,
\begin{equation} \label{eq:gb11exact}
-2\varepsilon = \mathcal{S}(r)+\nabla_r\cdot\boldsymbol{\mathcal{F}}(r),
\end{equation}
where $\mathcal{F}\propto\rho v^3$ is the energy flux as a function of length-scale $r$ and $\mathcal{S}$ is a new term that vanishes for incompressible turbulence and contains the contributions of $\nabla\cdot\vect{v}$. Using the general definition of the increment $\delta\xi=\xi(\vect{x}+\vect{r})-\xi(\vect{x})\equiv\xi^\prime-\xi$ of any given variable $\xi$ at position $\vect{x}$ and separated by a distance $r$, the exact expression for the new term $\mathcal{S}(r)$ is \citep{GaltierBanerjee2011,BanerjeeGaltier2013}
\begin{equation} \label{eq:gb11s}
\mathcal{S}(r) = \langle(\nabla\cdot\vect{v})^\prime (R-E)\rangle_x + \langle(\nabla\cdot\vect{v}) (R^\prime-E^\prime)\rangle_x,
\end{equation}
with $R=\rho(\vect{v}\cdot\vect{v}^\prime/2+e^\prime)$ and $E=\rho(\vect{v}\cdot\vect{v}/2+e)$, where $e=\cs^2\ln(\rho/\rho_0)$ and $\langle\dots\rangle_x$ denotes an average over all positions $\vect{x}$ in the turbulent flow.

Assuming isotropy (which is typically fulfilled, at least in a statistical sense) and integrating over a sphere with radius $r$, Equation~(\ref{eq:gb11exact}) can be written as
\begin{equation}
-\frac{2}{3}\varepsilon_\mathrm{eff}\,r = \mathcal{F}(r)
\end{equation}
with an effective dissipation rate
\begin{equation} \label{eq:vareps}
\varepsilon_\mathrm{eff}(r) = \varepsilon + \frac{3}{8} r \frac{\partial}{\partial r} \mathcal{S}|_{r\to0}
\end{equation}
to first order in a Taylor expansion of $\mathcal{S}$ for sufficiently small $r$, but still larger than the viscous scale to probe the scaling in the inertial range \citep[see Equation~15 in][]{GaltierBanerjee2011}.

Following dimensional analysis, the flux $\mathcal{F}\propto \rho v^3 \propto \varepsilon_\mathrm{eff}\,r$. Introducing again the density-weighted velocity $v_\mathrm{mw}\equiv\rhothreev$, we find $\varepsilon_\mathrm{eff}\,r \propto v_\mathrm{mw}^3$ and thus the spectrum of the density-weighted velocity
\begin{equation} \label{eq:gb11scaling}
P(\rhothreev) \propto \frac{\deriv v_\mathrm{mw}^2}{\deriv k} \propto \varepsilon_\mathrm{eff}^{2/3} r^{5/3} \propto \varepsilon_\mathrm{eff}^{2/3} k^{-5/3}.
\end{equation}
If $\varepsilon_\mathrm{eff} = \mathrm{const}$, then the spectrum $P(\rhothreev)\propto k^{-5/3}$ is expected to follow K41 scaling as argued in our previous derivation above (Equation~\ref{eq:k07scaling}). If, however, $\varepsilon_\mathrm{eff}$ scales with $r$ to some power, then $P(\rhothreev)$ does not follow $k^{-5/3}$, but is modified by the scaling of $\varepsilon_\mathrm{eff}(r)\propto \varepsilon + \mathcal{S}(r)$ according to Equation~(\ref{eq:vareps}). It is thus the additional term $\mathcal{S}$ in the derivation of \citet{GaltierBanerjee2011} that can lead to a modified scaling of $P(\rhothreev)$. Finally, \citet{GaltierBanerjee2011} argue that one may expect a scaling $\varepsilon_\mathrm{eff}\propto\mathcal{S}(r)\propto r^{2/3}$ for turbulence with a strong $\nabla\cdot\vect{v}$ component (see the dependence of $\mathcal{S}$ on $\nabla\cdot\vect{v}$ in Equation~\ref{eq:gb11s}), in which case we would obtain $P(\rhothreev)\propto k^{-19/9}$ according to Equation~(\ref{eq:gb11scaling}).

In the next section, we run and analyse two extremely high-resolution simulations of supersonic turbulence with solenoidal and compressive driving to test the prediction of $P(\rhothreev)\propto k^{-19/9}$ by \citet{GaltierBanerjee2011}. We do this here for compressive driving at a very high Mach number ($\mach=17$), such that $\nabla\cdot\vect{v}$ is potentially very strong on certain scales in the turbulent flow. We also analyse how $\nabla\cdot\vect{v}$ depends on the driving mode. A direct measurement of the new term $\mathcal{S}(r)$ is beyond the scope of this paper and will be presented in a future study with focus on the analysis of structure functions. Here we concentrate on the scaling inferred by Fourier analysis.

\begin{table*}
\caption{Simulation parameters and statistical measures}
\label{tab:sims}
\begin{tabular}{lrrrrrrrr}
\hline
Model & $\mach$ & Driving & $N_\mathrm{res}^3$ & PDF $\sigs$ & PDF $ \theta$ & Slope $P(v)$ & Slope $P(\rhothreev)$ & Slope $P(\nabla\cdot\vect{v})$ \\
(1) & (2) & (3) & (4) & (5) & (6) & (7) & (8) & (9)\\
\hline
M17sol256   & $17.2\pm1.0$ & Solenoidal & $256^3$   & $2.25\pm0.10$ & $0.37\pm0.06$ & n/a & n/a & n/a \\
M17sol512   & $17.1\pm0.9$ & Solenoidal & $512^3$   & $2.18\pm0.08$ & $0.28\pm0.04$ & n/a & n/a & n/a \\
M17sol1024 & $17.3\pm0.9$ & Solenoidal & $1024^3$ & $2.09\pm0.04$ & $0.22\pm0.02$ & n/a & n/a & n/a \\
M17sol2048 & $17.4\pm0.8$ & Solenoidal & $2048^3$ & $2.00\pm0.02$ & $0.18\pm0.02$ & n/a & n/a & n/a \\
M17sol4096 & $17.4\pm1.1$ & Solenoidal & $4096^3$ & $2.00\pm0.02$ & $0.20\pm0.02$ & $-1.96\pm0.04$ & $-1.74\pm0.05$ & $-0.08\pm0.05$ \\
\hline
M17comp256   & $16.6\pm1.0$ & Compressive & $256^3$   & $4.03\pm0.16$ & $0.60\pm0.08$ & n/a & n/a & n/a \\
M17comp512   & $16.9\pm1.1$ & Compressive & $512^3$   & $3.72\pm0.13$ & $0.43\pm0.07$ & n/a & n/a & n/a \\
M17comp1024 & $16.9\pm1.3$ & Compressive & $1024^3$ & $3.60\pm0.11$ & $0.39\pm0.06$ & n/a & n/a & n/a \\
M17comp2048 & $16.8\pm1.1$ & Compressive & $2048^3$ & $3.60\pm0.14$ & $0.39\pm0.07$ & n/a & n/a & n/a \\
M17comp4096 & $16.7\pm1.1$ & Compressive & $4096^3$ & $3.54\pm0.13$ & $0.37\pm0.06$ & $-1.99\pm0.03$ & $-2.10\pm0.07$ & $-0.00\pm0.03$ \\
\hline
\end{tabular}\\\vspace{0.05cm}
\raggedright{
\textbf{Notes.} Column (1): simulation name. Columns (2)--(4): rms Mach number, driving mode, and grid resolution. Columns (5) and (6): standard deviation of logarithmic density fluctuations $\sigs$ and the intermittency parameter $\theta$ for the density PDF fit. Columns (7)--(9): slopes of the Fourier power spectra for velocity, $\rhothreev$, and $\nabla\cdot\vect{v}$ (only measured with sufficient confidence for the $4096^3$ models).
}
\end{table*}

\section{Numerical approach} \label{sec:methods}

We use the \textsc{flash} code \citep{FryxellEtAl2000,DubeyEtAl2008} in its current version (v4) to solve the compressible gas-dynamical equations on three-dimensional, uniform, periodic grids of fixed side length $L$ with resolutions of $256^3$, $512^3$, $1024^3$, $2048^3$, and $4096^3$ grid points. To guarantee stability and accuracy of the numerical solution of the Euler equations, we use the HLL5R positive-definite Riemann solver \citep{WaaganFederrathKlingenberg2011}, closed with an isothermal equation of state, which is a reasonable approximation for dense, molecular gas of solar metallicity, over a wide range of densities \citep{OmukaiEtAl2005}. Keeping the gas temperature fixed has also the desirable advantage that the sound speed $\cs$ in the medium is fixed and thus the root-mean-square (rms) Mach number $\mach$ does not change systematically for a constant kinetic energy injection rate of the turbulence. This allows us to run these calculations for an arbitrary number of turbulent turnover times, $T=L/(2\cs\mach)$ \citep[following the definition by][]{KritsukEtAl2007,FederrathDuvalKlessenSchmidtMacLow2010}, to obtain a number of statistically independent flow snapshots, which can be averaged over time to yield converged statistical measures (PDFs and Fourier spectra).

To drive turbulence, we apply a stochastic acceleration field $\vect{F}_\mathrm{stir}$ as a momentum and energy source term. $\vect{F}_\mathrm{stir}$ only contains large-scale modes, $1<\left|\vect{k}\right|L/2\pi<3$, where most of the power is injected at the $\kinj=2$ mode in Fourier space, i.e., on half of the box size (for simplicity, we will drop the wavenumber unit $L/2\pi$ in the following). Such large-scale driving is favoured by molecular cloud observations \citep[e.g.,][]{OssenkopfMacLow2002,HeyerWilliamsBrunt2006,BruntHeyerMacLow2009,RomanDuvalEtAl2011}. The turbulence on smaller scales, $k\geq3$, is not directly affected by the driving and develops self-consistently there. We use the stochastic Ornstein-Uhlenbeck process to model $\vect{F}_\mathrm{stir}$ with a finite autocorrelation timescale \citep{EswaranPope1988,SchmidtHillebrandtNiemeyer2006}, set to the turbulent crossing time on the largest scales of the system, $T$ \citep[for details, see][]{SchmidtEtAl2009,FederrathDuvalKlessenSchmidtMacLow2010,KonstandinEtAl2012}.

We decompose the driving field into its solenoidal and compressive parts by applying a projection in Fourier space. In index notation, the projection operator reads $\mathcal{P}_{ij}^\zeta\,(\vect{k}) = \zeta\,\mathcal{P}_{ij}^\perp+(1-\zeta)\,\mathcal{P}_{ij}^\parallel = \zeta\,\delta_{ij}+(1-2\zeta)\,k_i k_j/|\vect{k}|^2$, where $\mathcal{P}_{ij}^\perp$ and $\mathcal{P}_{ij}^\parallel$ are the solenoidal and compressive projection operators. This projection allows us to construct a solenoidal (divergence-free) or compressive (curl-free) acceleration field by setting $\zeta=1$ or $\zeta=0$, respectively.

Our aim here is to study the regime of \emph{highly supersonic turbulence} such as in the interstellar medium, so we chose to drive the turbulence to $\mach\approx17$ for both extreme cases of driving (solenoidal and compressive), which means that all the scales resolved in our calculations are in the supersonic regime, i.e., above the sonic scale \citep{VazquezBallesterosKlessen2003,FederrathDuvalKlessenSchmidtMacLow2010}. Assuming a power-law velocity scaling of the turbulence, $v(\ell)\propto\ell^{\alpha}\propto k^{-\alpha}$, the sonic scale $\ks$ (where the scale-dependent Mach number, \mbox{$\mathscr{M}(\ell)\propto v(\ell)$} has dropped to unity) can be estimated as
\begin{equation} \label{eq:ks}
\ks/\kinj \approx (1/\mach)^{-1/\alpha},
\end{equation}
because the Mach number on the injection scale is roughly equal to the rms Mach number, $\mathscr{M}(L/2)\approx\mach$. With $\mach\approx17$, $\kinj=2$, and $\alpha\approx1/2$ \citep[the approximate velocity scaling for supersonic turbulence found in][and confirmed in Section~\ref{sec:spectra} below]{Burgers1948,KritsukEtAl2007,SchmidtFederrathKlessen2008,FederrathDuvalKlessenSchmidtMacLow2010}, this leads to $\ks\approx578$, which is in the dissipation range of the turbulence, even in our highest resolution runs with $4096^3$ points. Thus, any resolved scales in our calculations are in the truly supersonic regime of turbulence, so we can exclude any potential contamination of the inferred supersonic scaling exponents by a transition region to subsonic flow around the sonic scale, because that transition region is on much smaller scales than we analyse here. A list of all numerical models and parameters is provided in Table~\ref{tab:sims}.

\section{Results} \label{sec:results}

In the following, we will primarily focus on comparing two simulations with solenoidal and compressive driving, each with a grid resolution of $4096^3$ points, which is currently the world's largest data set of supersonic turbulence \citep[an equivalent resolution was so far only reached for incompressible turbulence by][]{KanedaEtAl2003}. Each simulation was run for about 44,000 time steps (see Appendix~\ref{app:vpdfs}) on 32,768 compute cores running in parallel on SuperMUC at the Leibniz Rechenzentrum in Garching (which consumed about 7.2 million CPU hours altogether). Each run produced $115\,\mathrm{TB}$ of data (51 double-precision snapshots of the turbulent density and three-dimensional velocity, stretched over six turbulent turnover times). In order to study resolution effects, we also compare each $4096^3$ model with the respective lower-resolution versions with $2048^3$, $1024^3$, $512^3$, and $256^3$ compute cells (see Table~\ref{tab:sims} for a complete list of simulations).

\begin{figure*}
\centerline{\includegraphics[width=0.9\linewidth]{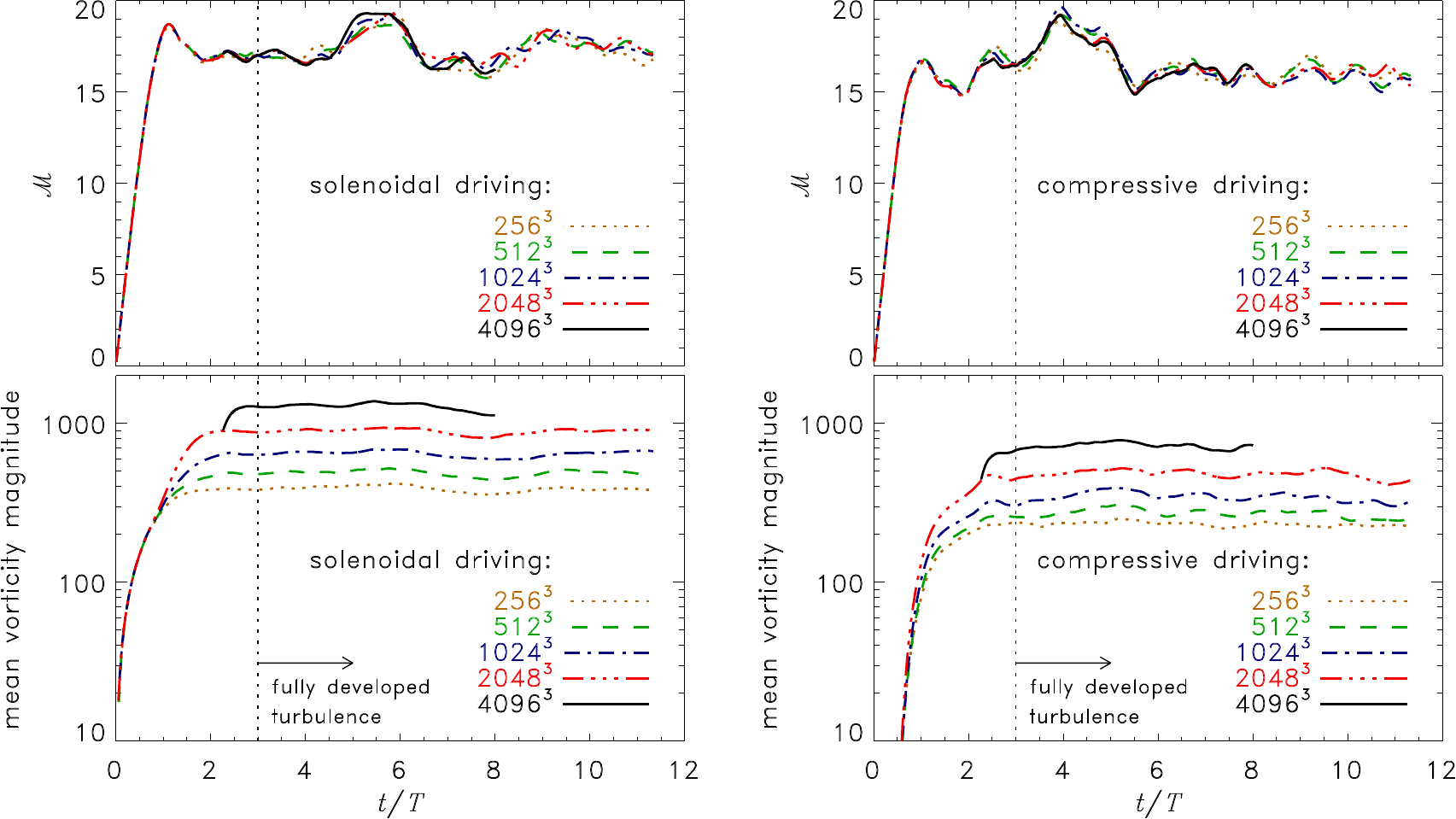}}
\caption{rms Mach number (top panels) and mean vorticity magnitude (bottom panels) as a function of time for solenoidal driving (left-hand column) and compressive driving (right-hand column). Different line styles indicate grid resolutions from $256^3$ to $4096^3$ cells. Both the Mach number and vorticity reach a statistically steady state at $t=2T$. The $4096^3$ calculations were initialized with the density and velocity fields of the $2048^3$ runs at $t=2T$. They reach statistical convergence within another turnover time and were run until $t=8T$. We thus use the interval $3 \leq t/T \leq 8$ for averaging in subsequent analyses (for all resolutions). The Mach number is well converged, while the vorticity increases with resolution as expected \citep[][]{Lesieur1997,SytineEtAl2000}. The vorticity at a fixed resolution is about a factor of 1.8 higher for solenoidal compared to compressive driving, consistent with the limit for hypersonic turbulence (factor of two) estimated in \citet{FederrathEtAl2011PRL}.}
\label{fig:tevol}
\end{figure*}

\subsection{Time evolution and turbulent structure}

For all but the $4096^3$ simulations, we start with gas of initial velocity $\vect{v}_0=0$ and homogeneous density $\rho_0$ in a three-dimensional periodic box. The driving then accelerates the gas to our target Mach number, $\mach\approx17$, until a statistically converged regime of fully developed turbulent flow is reached, which happens after about two turnover times, $2T$. For the $4096^3$ runs, we take the density and velocity fields of the respective $2048^3$ simulations at $t=2T$ and map them on $4096^3$ grids to spare the initial transient start-up phase, $t<2T$. We run them until $t=8T$, which gives us a sufficiently large statistical sample of independent flow snapshots to obtain converged results. In order to allow the turbulence to adjust to the new resolution and to converge to a statistically steady state, we start analysing the results for $t\geq3T$, leaving us five turnover times ($3\leq t/T\leq8$) to average PDFs and Fourier spectra. This procedure also allows us to quantify the temporal variations of the turbulence in the fully developed regime.

To demonstrate statistical convergence within \mbox{$3\leq t/T\leq8$}, we show the time evolution of the rms Mach number and the mean vorticity magnitude, $\langle\left|\nabla\times\vect{v}\right|\rangle$, in Figure~\ref{fig:tevol} for all resolutions with solenoidal driving (left-hand panels) and with compressive driving (right-hand panels). We see that both $\mach$ and the vorticity grow quickly within $2T$ and then reach a statistically steady state. The $4096^3$ runs, which were initialized with the $2048^3$ density and velocity fields at $t=2T$, reach a steady state by $t=3T$, so we choose to start averaging PDFs and spectra for $t\geq3T$, when all statistics have safely reached a steady state. (We also inspected the time evolution of the rms velocity divergence, as well as the time evolution of Fourier spectra shown in Section~\ref{sec:spectra}, all of which were statistically converged for $t\geq3T$.)

Figure~\ref{fig:tevol} shows that both extreme types of driving generate vorticity, with solenoidal driving being about twice as efficient. This is because solenoidal motions are directly injected by solenoidal driving, while they have to self-generate in shock collisions and by viscous interactions across density gradients with subsequent amplification in the case of purely compressive driving \citep[for details of the `anti-diffusion' term responsible for this behaviour, see the vorticity equations in][]{MeeBrandenburg2006,FederrathEtAl2011PRL}.

Since the overall vorticity is always dominated by small-scale structures, which have the smallest turbulent time-scales, $t(\ell)=\ell/v(\ell)\propto\ell^{1-\alpha}$ for any $0<\alpha<1$ (e.g., $\alpha=1/3$ for Kolmogorov and $\alpha=1/2$ for Burgers turbulence as reasonable limiting cases), increasing the resolution leads to higher levels of vorticity, as the effective viscosity of the gas decreases and the effective Reynolds number increases \citep[][]{SurEtAl2010,FederrathSurSchleicherBanerjeeKlessen2011}. This is consistent with the expectation that the vorticity tends to infinity at a finite time in the limit of zero viscosity \citep{Lesieur1997,SytineEtAl2000}. The effective Reynolds numbers of our simulations are of the order of $\mathrm{Re}\approx N_\mathrm{res}^{1+\alpha}$ \citep{BenziEtAl1993,FederrathEtAl2011PRL}. For Burgers turbulence with $\alpha=1/2$ and for a grid resolution of $N_\mathrm{res}=4096$, this yields $\mathrm{Re}\approx3\times10^5$. It must be emphasized though that the actual dissipation range of the turbulence is not resolved when computing numerical solutions of the Euler equations instead of the Navier-Stokes equations \citep{SytineEtAl2000}.  To find the trustworthy scales in our simulations, i.e., the inertial range, we have to study the resolution dependence of Fourier spectra, which we do below in Section~\ref{sec:spectra}.

\begin{figure*}
\centerline{\includegraphics[width=0.98\linewidth]{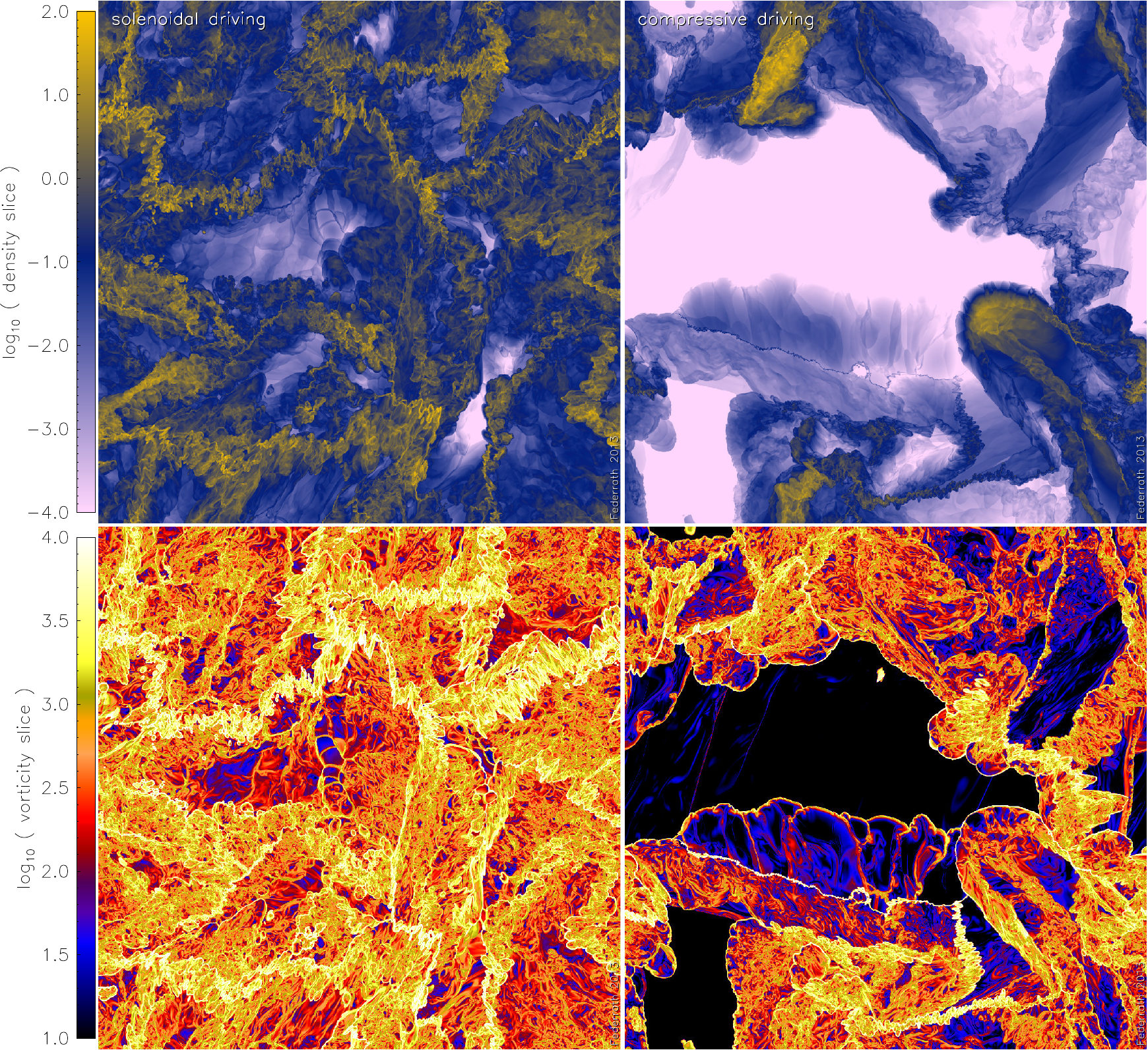}}
\caption{Slices through the three-dimensional gas density (top panels) and vorticity (bottom panels) for fully developed, highly compressible, supersonic turbulence, generated by solenoidal driving (left-hand column) and compressive driving (right-hand column), and a grid resolution of $4096^3$ cells. Large regions of very low density and very low vorticity in the compressive driving case indicate that the inertial range is shifted to slightly smaller scales for compressive driving compared to the more space filling case of solenoidal driving. The fractal dimension of the density is $D_\mathrm{f}\approx2.6$ and $D_\mathrm{f}\approx2.3$ for solenoidal and compressive driving, respectively \citep{FederrathKlessenSchmidt2009}. (Movies are available in the online version.)}
\label{fig:images}
\end{figure*}

\begin{figure*}
\centerline{\includegraphics[width=0.98\linewidth]{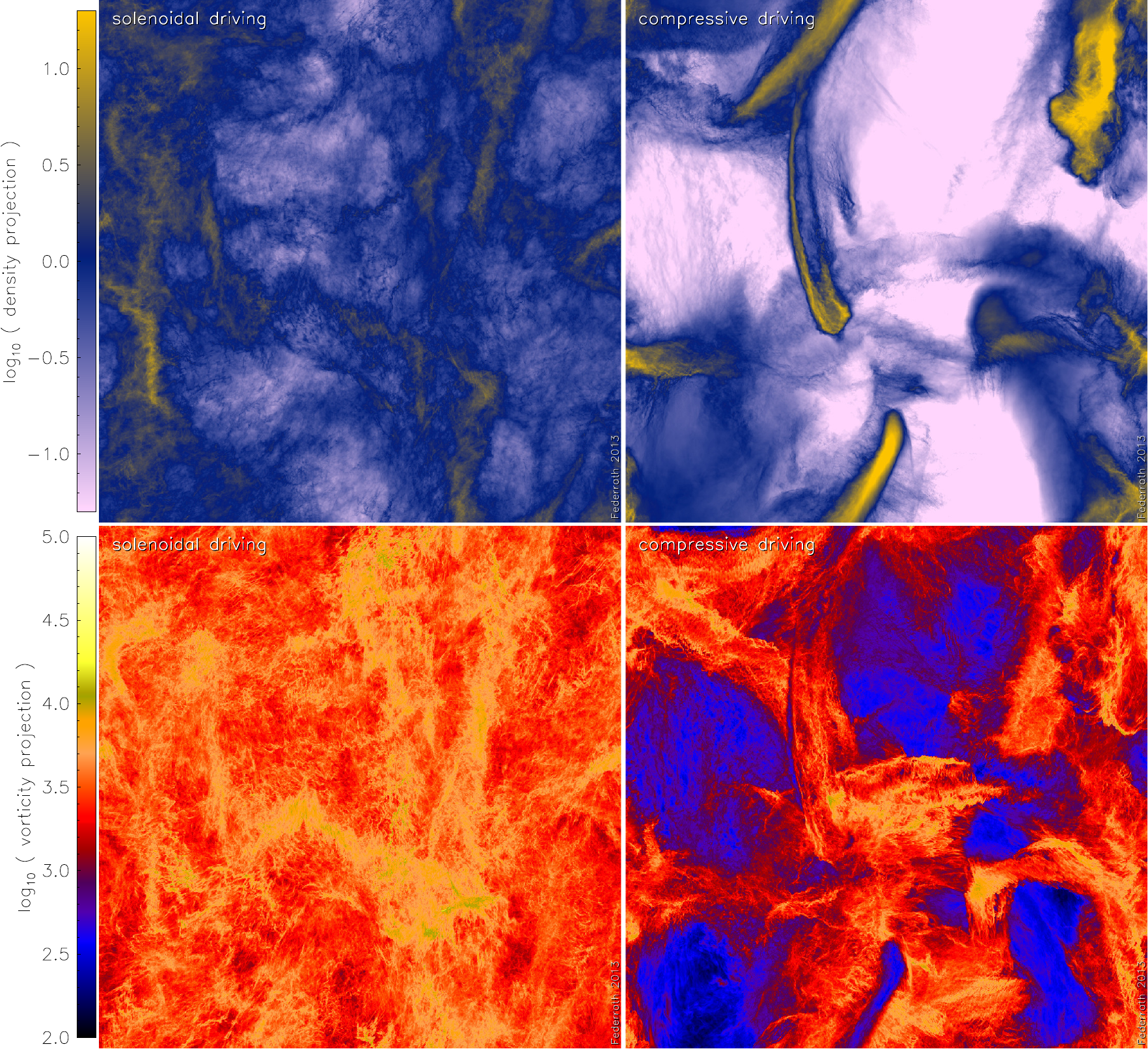}}
\caption{Same as Figure~\ref{fig:images}, but instead of slices, these show projections (integration along the line of sight) of the three-dimensional gas density (top panels) and (mass-weighted) vorticity (bottom panels) for solenoidal driving (left-hand column) and compressive driving (right-hand column). (Movies are available in the online version.)}
\label{fig:images_proj}
\end{figure*}

Slices through the three-dimensional turbulent flow structures are shown in Figure~\ref{fig:images}. Corresponding projections (integration along the line of sight) of the density and the mass-weighted vorticity are shown in Figure~\ref{fig:images_proj}. The latter is closer to what an observer would see in a molecular cloud observation. The gas density and vorticity appear to be correlated for both driving types in the slices and in the projections. This is because vorticity is primarily generated in shocks and across strong density gradients \citep{MeeBrandenburg2006,FederrathEtAl2011PRL}. Solenoidal driving produces more space-filling structures with a fractal dimension $D_\mathrm{f}\approx2.6$, while $D_\mathrm{f}\approx2.3$ and thus closer to sheets for purely compressive driving \citep{FederrathKlessenSchmidt2009}. The latter is consistent with the fractal mass dimension inferred for molecular clouds in the Milky Way \citep{RomanDuvalEtAl2010} and in nearby galaxies \citep{DonovanMeyerEtAl2013}. Dense structures with high levels of vorticity are confined to relatively small patches in the case of purely compressive driving. Some large-scale regions with sizes of about $\ell\gtrsim L/10$ or $k\lesssim10$ remain almost empty and exhibit very low gas density and vorticity. However, structures on smaller scales ($k\gtrsim10$) do show high levels of vorticity throughout. This indicates that the inertial range in $\mach\approx17$ turbulence with compressive driving starts on somewhat smaller scales than with purely solenoidal driving \citep[][]{KritsukEtAl2010}, which is different from the case of mildly supersonic turbulence with \mbox{$\mach\approx5$--$6$} \citep{KritsukEtAl2007,FederrathDuvalKlessenSchmidtMacLow2010}, where the inertial-range extent is not significantly different between solenoidal and compressive driving.

\begin{figure*}
\centerline{\includegraphics[width=0.90\linewidth]{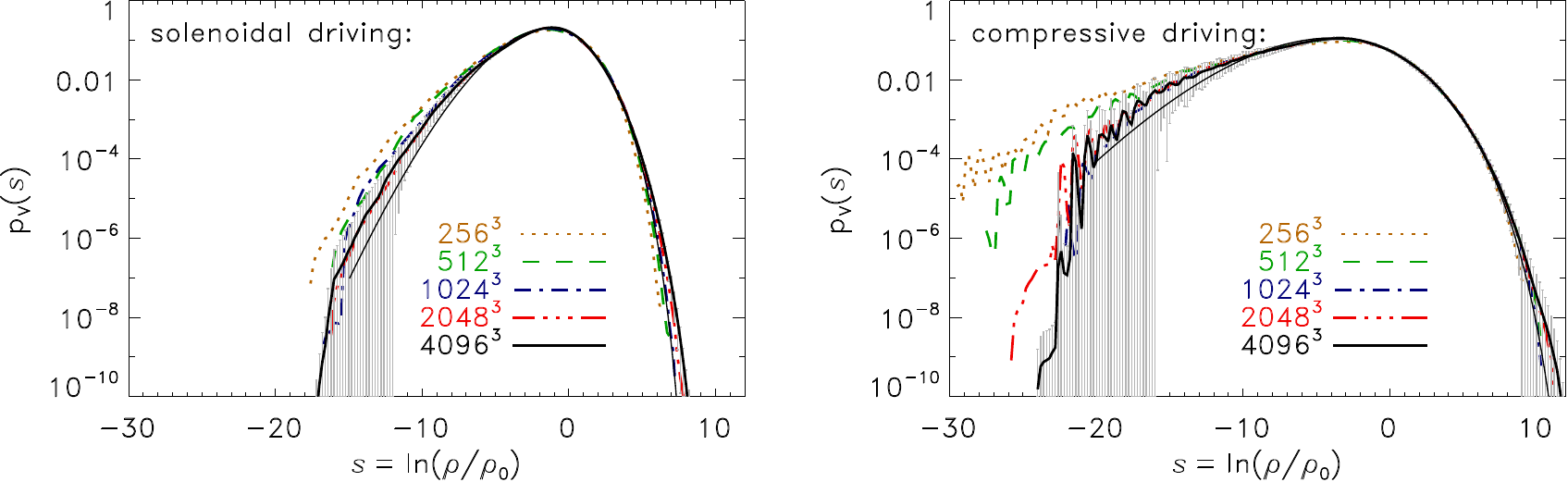}}
\caption{Volume-weighted PDFs of the logarithmic gas density $s\equiv\ln(\rho/\rho_0)$ for solenoidal driving (left-hand column) and compressive driving (right-hand column). Numerical resolutions from $256^3$ to $4096^3$ are plotted with different line styles according to the legend. The grey error bars indicate the $1\sigma$ snapshot-to-snapshot variations (only shown for the $4096^3$ runs). The thin black lines show two-parameter fits for $\sigs$ and $\theta$ to the $4096^3$ data with the intermittency PDF, Equation~(\ref{eq:pdffit}). Compressive driving produces a larger standard deviation $\sigs$ than solenoidal driving, and exhibits a much higher degree of intermittency (quantified by the fit parameter $\theta$; see Table~\ref{tab:sims} and Figure~\ref{fig:pdfconv}).}
\label{fig:pdfs}
\end{figure*}

\subsection{Density PDFs}

The strong density variations in supersonic turbulence, such as seen in Figures~\ref{fig:images} and~\ref{fig:images_proj}, are clearly the most prominent difference to incompressible turbulence. To quantify these, we briefly analyse the PDF of the gas density. The volume-weighted density PDFs of the logarithmic density $s\equiv\ln(\rho/\rho_0)$ are shown in Figure~\ref{fig:pdfs}. Obviously, compressive driving produces a significantly wider density distribution with a larger standard deviation than solenoidal driving for the same rms Mach number, which has been explored and discussed in detail in previous studies \citep{FederrathKlessenSchmidt2008,PriceFederrathBrunt2011,KonstandinEtAl2012,KonstandinEtAl2012ApJ}.

Previous works suggest that the density PDF should be approximately log-normal \citep{Vazquez1994,PadoanNordlundJones1997,PassotVazquez1998}. Deviations from perfectly log-normal distributions are caused by intermittency and sampling effects \citep{KritsukEtAl2007,KowalLazarianBeresnyak2007,FederrathDuvalKlessenSchmidtMacLow2010,PriceFederrath2010,KonstandinEtAl2012ApJ}. Recently, \citet{Hopkins2013PDF} suggested an intermittency fit for the volume-weighted PDF with the following function,

\begin{eqnarray} \label{eq:pdffit}
p_V(s) = I_1\left(2\sqrt{\lambda\,\omega(s)}\right)\exp\left[-\left(\lambda+\omega(s)\right)\right]\sqrt{\frac{\lambda}{\theta^2\,\omega(s)}}\,,\nonumber\\
\lambda\equiv\sigs^2/(2 \theta^2),\quad\omega(s)\equiv\lambda/(1+ \theta)-s/ \theta \;\; (\omega\geq0)
\end{eqnarray}
where $I_1(x)$ is the modified Bessel function of the first kind. Equation~(\ref{eq:pdffit}) is motivated and explained in detail in \citet{Hopkins2013PDF}. It contains two parameters, the standard deviation of logarithmic density variations, $\sigs$, and an intermittency parameter $ \theta$. In the zero-intermittency limit $ \theta\to0$, Equation~(\ref{eq:pdffit}) simplifies to a log-normal PDF. \citet{Hopkins2013PDF} show that this intermittency form of the PDF provides excellent fits to density PDFs from turbulence simulations with extremely different properties (solenoidal, mixed, and compressive driving, Mach numbers from 0.1 to 18, and varying degrees of magnetization).

\begin{figure*}
\centerline{\includegraphics[width=0.90\linewidth]{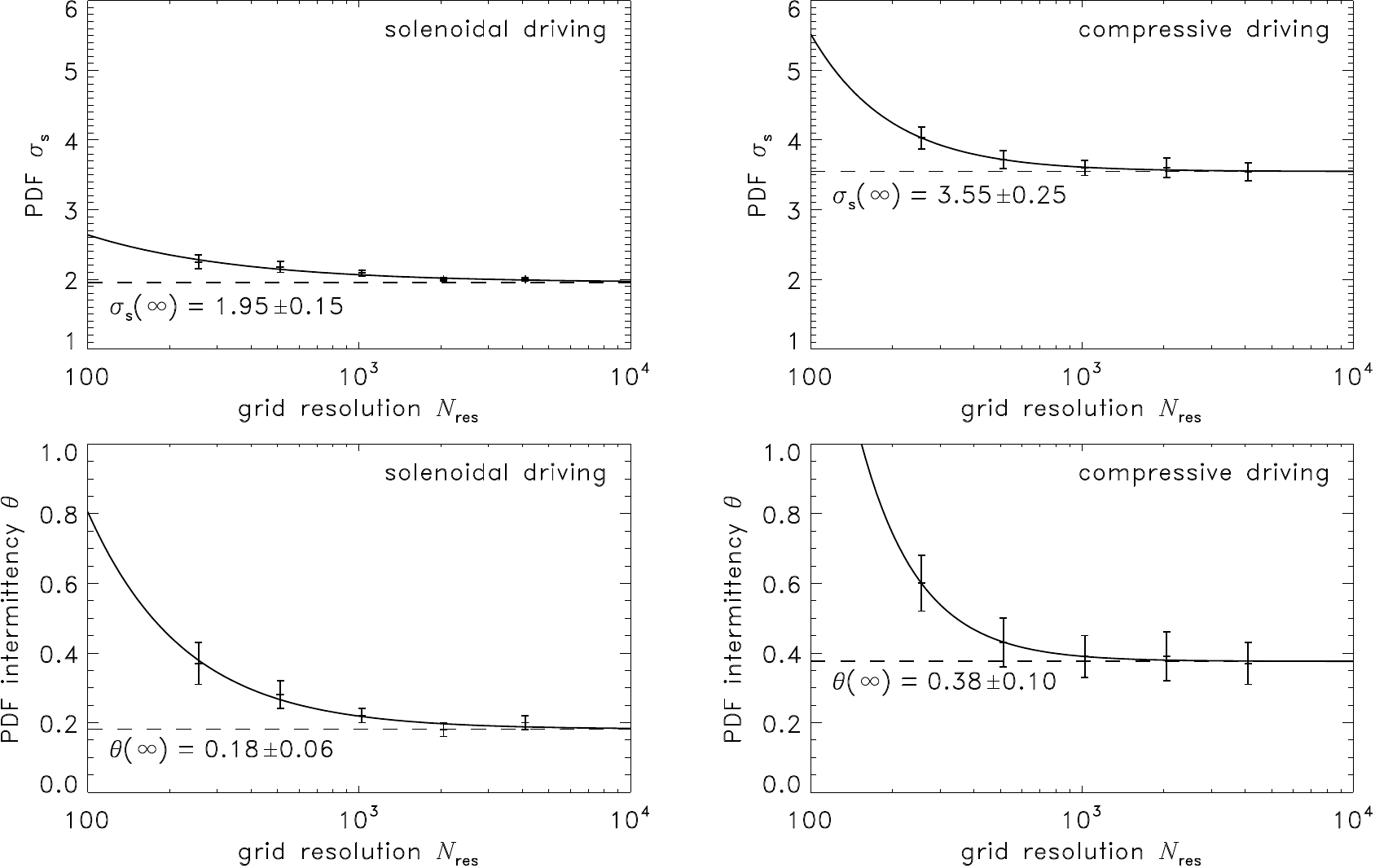}}
\caption{Numerical convergence study of the PDF standard deviation $\sigs$ (top panels) and intermittency $\theta$ (bottom panels) for solenoidal driving (left-hand column) and compressive driving (right-hand column). The solid line is a power-law convergence fit with $y(x)=ax^{-b}+c$ and the dashed line shows the limit of infinite resolution (fit parameter $c$). The $4096^3$ runs are almost converged to the infinite-resolution limit. Compressive driving is characterized by significantly larger $\sigs$ and stronger intermittency $\theta$ compared to solenoidal driving.}
\label{fig:pdfconv}
\end{figure*}

We apply fits to all PDFs in Figure~\ref{fig:pdfs} for different driving and resolutions by simultaneously fitting the standard deviation $\sigs$ and the intermittency parameter $\theta$, i.e., we perform a two-parameter fit. We note that this yields fitted values of $\sigs$ that agree very well with the actual data values (to within 10\%). The parameters $\sigs$ and $ \theta$ are listed in Table~\ref{tab:sims}. Comparing different resolutions, we see that $\sigs$ and $ \theta$ decrease with increasing resolution. To study the convergence behaviour, we plot $\sigs$ and $ \theta$ as a function of resolution in Figure~\ref{fig:pdfconv}. The top panels show $\sigs$ and the bottom panels show $\theta$. We apply power-law fits with the model function $y(x)=ax^{-b}+c$ to study convergence and to estimate the parameter values $c$, which correspond to the limit of infinite numerical resolution $N_\mathrm{res}\to\infty$. We perform fits for $c=\sigs(\infty)$ and $c=\theta(\infty)$ and both driving types. The fit curves are added in each panel of Figure~\ref{fig:pdfconv}. They fit the data for all our resolutions \mbox{$N_\mathrm{res}=256$--$4096$} quite well and give $\sigs(\infty)=1.95\pm0.15$ and $\theta(\infty)=0.18\pm0.06$ for solenoidal driving, and $\sigs(\infty)=3.55\pm0.25$ and $\theta(\infty)=0.38\pm0.10$ for compressive driving in the limit of infinite resolution. The $2048^3$ and $4096^3$ data are converged to within 10\% of the limit $N_\mathrm{res}\to\infty$.

Our fit values for the infinite-resolution limit in Figure~\ref{fig:pdfconv} show that compressive driving is significantly more intermittent with $\theta_\mathrm{comp}/\theta_\mathrm{sol}\approx2.1$, consistent with the fits in \citet{Hopkins2013PDF} for $\mach\approx15$ simulations with solenoidal and compressive driving by \citet{KonstandinEtAl2012ApJ}.

\begin{figure*}
\centerline{\includegraphics[width=0.95\linewidth]{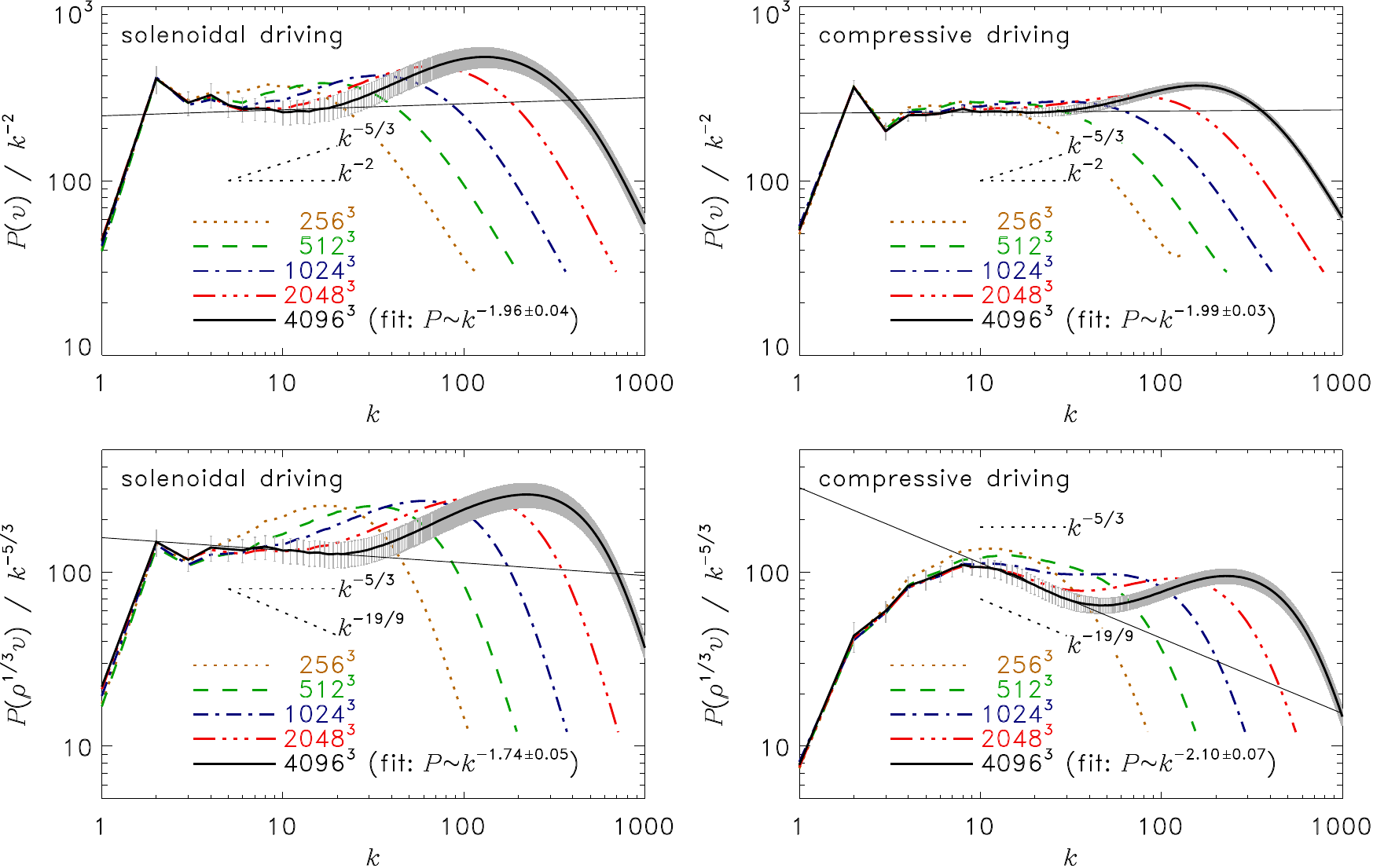}}
\caption{Compensated Fourier power spectra of the velocity, $P(v) / k^{-2}$ (top panels), and the density-weighted velocity, $P(\rhothreev) / k^{-5/3}$ (bottom panels) for solenoidal driving (left-hand column) and compressive driving (right-hand column). Different line styles show different grid resolutions and the grey error bars indicate the $1\sigma$ temporal variations (only shown for the $4096^3$ data). The extent of the scaling range (inertial range) is indicated by the dotted lines, showing different power-law scalings in each panel, for comparison. The thin solid lines in each panel are power-law fits within the most reasonable scaling ranges ($5\leq k \leq 20$ for solenoidal driving and $10 \leq k \leq 30$ for compressive driving), considering how each of the spectra changes with increasing resolution and considering contamination by the bottleneck effect (see text for details).}
\label{fig:spect}
\end{figure*}

\subsection{Fourier power spectra of $v$ and $\rhothreev$} \label{sec:spectra}

Fourier power spectra are an ideal tool to study the scaling of fluid variables such as the turbulent velocity, density, or combinations of both, and the results are readily comparable to turbulence theories such as the incompressible \citet[][hereafter K41]{Kolmogorov1941c} model or the \citet[][hereafter B48]{Burgers1948} model. The latter is entirely composed of discontinuities (or shocks). As the analysis is done in Fourier space, the spatial scale $\ell$ simply transforms to a wavenumber scale $k=2\pi/\ell$. The three-dimensional Fourier transform of a variable $q(\boldsymbol{\ell})$ with $\boldsymbol{\ell}=\{\ell_1,\ell_2,\ell_3\}$ is defined as
\begin{equation} \label{eq:ft}
\widehat{q}(\boldsymbol{k}) = \frac{1}{(2\pi L)^{3/2}} \int q(\boldsymbol{\ell})\,e^{-i\,\boldsymbol{k}\cdot\boldsymbol{\ell}}\,\deriv^3 \ell\,,
\end{equation}
where we denote the Fourier transform of $q(\boldsymbol{\ell})$ with $\widehat{q}(\boldsymbol{k})$. With this definition of the Fourier transform, the Fourier power spectrum of $q$ is given by
\begin{equation} \label{eq:pthreed}
P(q) = \langle \widehat{q}\cdot\widehat{q}^\star\,4\pi k^2 \rangle_k\,,
\end{equation}
as an average of $\widehat{q}\cdot\widehat{q}^\star$ over a spherical shell with radius $k=|\boldsymbol{k}|$ and thickness $\deriv k$ in Fourier space.

An important caveat of numerical turbulence simulations is that the Fourier spectra are typically only converged within a very tiny range of scales for the resolutions achievable with current technology \citep[see also][]{KleinEtAl2007}. Thus, a large fraction of scales is either affected by numerical dissipation (and thus not converged) or a potential inertial range is contaminated by the so-called bottleneck effect \citep{Falkovich1994,DoblerEtAl2003,SchmidtHillebrandtNiemeyer2006}. Scales affected by numerical dissipation and the bottleneck effect must be excluded from the analysis. Previous simulations established very stringent requirements on the numerical resolution. For instance, \citet{KritsukEtAl2007}, \citet{SchmidtEtAl2009}, \citet{LemasterStone2009}, and \citet{FederrathDuvalKlessenSchmidtMacLow2010} find that at least a resolution of $512^3$ grid cells is required, but even with a resolution of $1024^3$ grid cells, the scaling range is much less than half a decade. Those studies were also run at relatively low Mach number \mbox{($\mach\approx5$--$7$)}, while the larger Mach numbers studied here may require even higher resolution.

\citet{FederrathDuvalKlessenSchmidtMacLow2010,FederrathSurSchleicherBanerjeeKlessen2011} found a strict lower limit of 32 grid cells, across which the energy carried by a vortex is reasonably well captured in a grid-based code. Vortices resolved with less than 32 grid cells in diameter suffer numerical dissipation. It is quite obvious that vortices are completely lost when their diameter falls below a single grid cell. In addition to that, the bottleneck effect can contaminate the inertial-range scaling on even larger scales than numerical dissipation. Eventually, only resolution studies can reveal the trustworthy scales, which is why we perform a resolution study below. Here we compare runs with $256^3$, $512^3$, $1024^3$, $2048^3$, and $4096^3$ grid cells, showing that the most reasonable scaling ranges for the simulations studied here are $5\leq k \leq 20$ for solenoidal driving and $10 \leq k \leq 30$ for compressive driving. The upper limit for solenoidal driving extends to slightly lower $k$ than with compressive driving (i.e., $k_\mathrm{max}=20$ versus $30$), because the bottleneck effect is stronger for solenoidal driving (Konstandin et al.~2013, in preparation). On the other hand, the lower limit ($k_\mathrm{min}=5$ versus $10$) is shifted to higher $k$ for compressive driving, as we guessed from the visual inspection of Figures~\ref{fig:images} and~\ref{fig:images_proj} (right-hand panels), which showed large empty patches of size down to one-tenth of the box length, i.e., $k_\mathrm{min}\approx10$.

Figure~\ref{fig:spect} (top panels) shows the compensated velocity spectra, $P(v)$, i.e., setting $q\equiv\vect{v}$ (the turbulent velocity) in Equation~(\ref{eq:pthreed}). We see that both solenoidal and compressive driving produce velocity scalings much steeper than the K41 scaling for incompressible turbulence ($P\propto k^{-5/3}$). Solenoidal driving yields $P(v)\propto k^{-1.96\pm0.04}$ and compressive driving yields $P(v)\propto k^{-1.99\pm0.03}$, both very close to B48 scaling ($P\propto k^{-2}$). These results for $P(v)$ are consistent with previous studies by \citet{KritsukEtAl2007}, \citet{LemasterStone2009}, and \citet{FederrathDuvalKlessenSchmidtMacLow2010} at a lower Mach number (\mbox{$\mach\approx5$--$7$}) and lower resolution. Moreover, our measured spectral slopes for solenoidal and compressive driving are both consistent with a turbulent velocity dispersion--size scaling \citep[often referred to as][relation]{Larson1981} of $v\propto\ell^{0.5}$, as measured in observational studies of molecular clouds \citep{Larson1981,SolomonEtAl1987,FalgaronePugetPerault1992,OssenkopfMacLow2002,HeyerBrunt2004,RomanDuvalEtAl2011}.

In contrast to the pure velocity spectra, Figure~\ref{fig:spect} (bottom panels) shows that the density-weighted velocity spectra $P(\rhothreev)$ are significantly different for different driving. As explained in Section~\ref{sec:theory}, the density-weighted velocity $\rhothreev$ has been proposed to exhibit a more universal scaling in supersonic turbulence than the pure velocity. According to the simple theoretical analysis given by Equation~(\ref{eq:k07scaling}), we would expect $P(\rhothreev)\propto k^{-5/3}$, as in the incompressible K41 case. However, we see that contrary to the hypothesis of universality of $P(\rhothreev)$, the spectra are significantly different between solenoidal and compressive driving. While solenoidal driving is close to (but slightly steeper than) K41 scaling with $k^{-1.74\pm0.05}$, compressive driving exhibits a significantly steeper scaling with $k^{-2.10\pm0.07}$. The latter seems to be consistent with the recent theoretical prediction of $P(\rhothreev)\propto k^{-19/9}$ by \citet{GaltierBanerjee2011} for highly compressible turbulence with a strong $\nabla\cdot\vect{v}$ component, which we discuss further in Section~\ref{sec:why}.

Our resolution study in Figure~\ref{fig:spect} shows that the inertial range is shifted to smaller scales for compressive driving, as we guessed from the visual inspection of Figures~\ref{fig:images} and~\ref{fig:images_proj} (right-hand panels), showing large empty patches of size down to one-tenth of the box length, i.e., $k\approx10$. This was not the case in our previous simulations with compressive driving and a moderate Mach number of \mbox{$\mach\approx5$--$6$} in \citet{FederrathDuvalKlessenSchmidtMacLow2010}, which were consistent with $P(\rhothreev)\propto k^{-19/9}$ at moderate \mbox{$512^3$--$1024^3$} resolution. Resolving Mach 17 turbulence requires much higher resolution than Mach 6 turbulence. The shock width decreases with the Mach number squared. Thus, we have to resolve structures that are $(17/6)^2 \approx 8$ times smaller compared to \citet{FederrathDuvalKlessenSchmidtMacLow2010}, which is about the difference in the required resolution, i.e., $4096/512$ is a factor of 8 in linear resolution, equivalent to the reduction in shock width between Mach 6 and Mach 17 turbulence.

From this analysis, we see that at least a resolution of $4096^3$ grid cells is required to resolve the inertial range for supersonic turbulence with a higher Mach number ($\mach\approx17$). Even with such high resolution, the inertial range only extends between $k\approx10$ and $30$. We emphasize that the $P(\rhothreev)$ spectrum for compressive driving would have been consistent with the universal hypothesis of $k^{-5/3}$ scaling, if we had only resolved it up to $1024^3$ grid cells (see the relatively flat dashed and dot-dashed lines for $512^3$ and $1024^3$ resolutions in the bottom, right-hand panel of Figure~\ref{fig:spect}). In contrast, the additional $2048^3$ and $4096^3$ calculations clearly demonstrate a significant steepening to $P(\rhothreev)\propto k^{-19/9}$ in the scaling range $10 \leq k \leq 30$, before $P(\rhothreev)$ flattens again for $k\gtrsim40$ due to bottleneck contamination there. The steepening to $k^{-19/9}$ is basically absent for resolutions $N_\mathrm{res}\lesssim1024$, because the bottleneck effect artificially flattens the spectra, i.e., we would have measured a shallower slope closer to $k^{-5/3}$, if we had included scales in the fit that are affected by the bottleneck effect. Thus, for fitting the spectra, great care must be exercised in choosing converged scales that reflect the physical scaling of supersonic turbulence, which requires extremely high resolution for $\mach\gtrsim15$ turbulence, such as in many molecular clouds.

Given all statistical and numerical uncertainties, and given the systematic evolution of the spectra with increasing resolution in Figure~\ref{fig:spect}, our measured slopes for the $4096^3$ models are converged to within an uncertainty of $<10\%$, which we estimated by extrapolating the slopes for $1024^3$, $2048^3$, and $4096^3$ resolutions to the limit of infinite resolution (the temporal variations are of the order of $<5\%$). Varying the fit range arbitrarily between $k_\mathrm{min}=5$ and $k_\mathrm{max}=30$ (given the constraint that $k_\mathrm{max}/k_\mathrm{min}\geq2$) changes the measured slopes by less than $10\%$. Thus, the $P(\rhothreev)$ slopes are significantly different between solenoidal and compressive driving.

\begin{figure*}
\centerline{\includegraphics[width=0.95\linewidth]{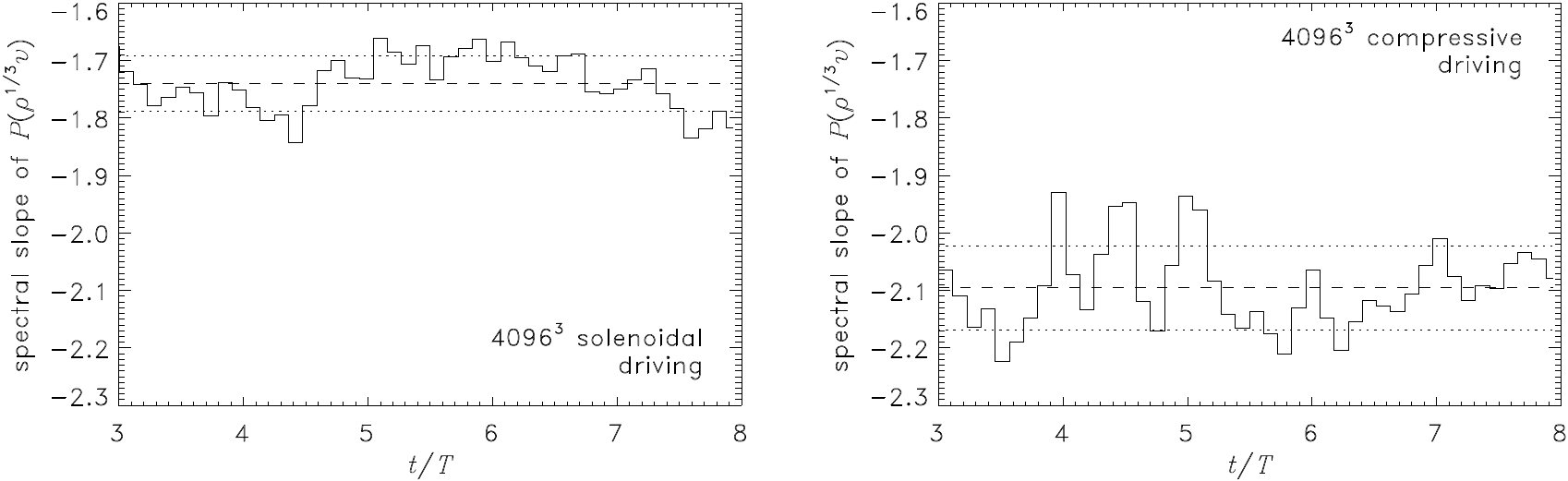}}
\caption{Spectral slope of the density-weighted velocity power spectra $P(\rhothreev)$ in the regime of fully developed turbulence, \mbox{$3\leq t/T\leq8$}. The left panel is for solenoidal driving and the right panel for compressive driving. The dashed line shows the time-averaged mean slope and the dotted lines enclose the $1\sigma$ time variations. Solenoidal driving yields a slope of $-1.74\pm0.05$, close to K41 scaling, while compressive driving gives a significantly steeper slope of $-2.10\pm0.07$, consistent with the theoretical prediction in \citet{GaltierBanerjee2011}.
}
\label{fig:spectevol}
\end{figure*}

To see that the slopes of the $P(\rhothreev)$ spectra are also converged in time, we fit each individual flow snapshot within the fully developed regime of turbulence. This analysis is shown in Figure~\ref{fig:spectevol}, demonstrating convergence and emphasizing our main result: the spectral slope of the density-weighted velocity $\rhothreev$ is $-1.74\pm0.05$, only slightly steeper than K41 scaling, while compressive driving yields a significantly steeper slope of $-2.10\pm0.07$, consistent with the theoretical prediction in \citet{GaltierBanerjee2011}.

\subsection{Why is the scaling of $\rhothreev$ not universal?} \label{sec:why}

The reason for the dependence of $P(\rhothreev)$ on the driving that we found above can be seen in the theoretical derivation by \citet{GaltierBanerjee2011} of the scaling in Equation~(\ref{eq:gb11scaling}). This derivation shows that $P(\rhothreev)\propto \varepsilon_\mathrm{eff}^{2/3} k^{-5/3}$. Since $\varepsilon_\mathrm{eff}(r)\propto \mathcal{S}(r)$, exactly defined in Equation~(\ref{eq:gb11s}), $\varepsilon_\mathrm{eff}$ is not constant, but instead modifies the scaling of $P(\rhothreev)$. Preliminary analyses of structure functions (not shown here) indicate a positive power-law scaling of $\mathcal{S}(r)$, such that the effect of the new (compressible) term $\mathcal{S}(r)$ is to \emph{steepen} the slope of $P(\rhothreev)$ compared to K41 scaling with $k^{-5/3}$. Such a steepening is indeed seen in the numerical spectra in Figure~\ref{fig:spect} with $P(\rhothreev)\propto k^{-1.74}$ and $\propto k^{-2.10}$ for solenoidal and compressive driving, respectively. The stronger steepening for compressive driving is caused by the stronger $\nabla\cdot\vect{v}$ component for this type of driving (see Equation~\ref{eq:gb11s} for the dependence of $\mathcal{S}$ on $\nabla\cdot\vect{v}$).

\begin{figure*}
\centerline{\includegraphics[width=0.95\linewidth]{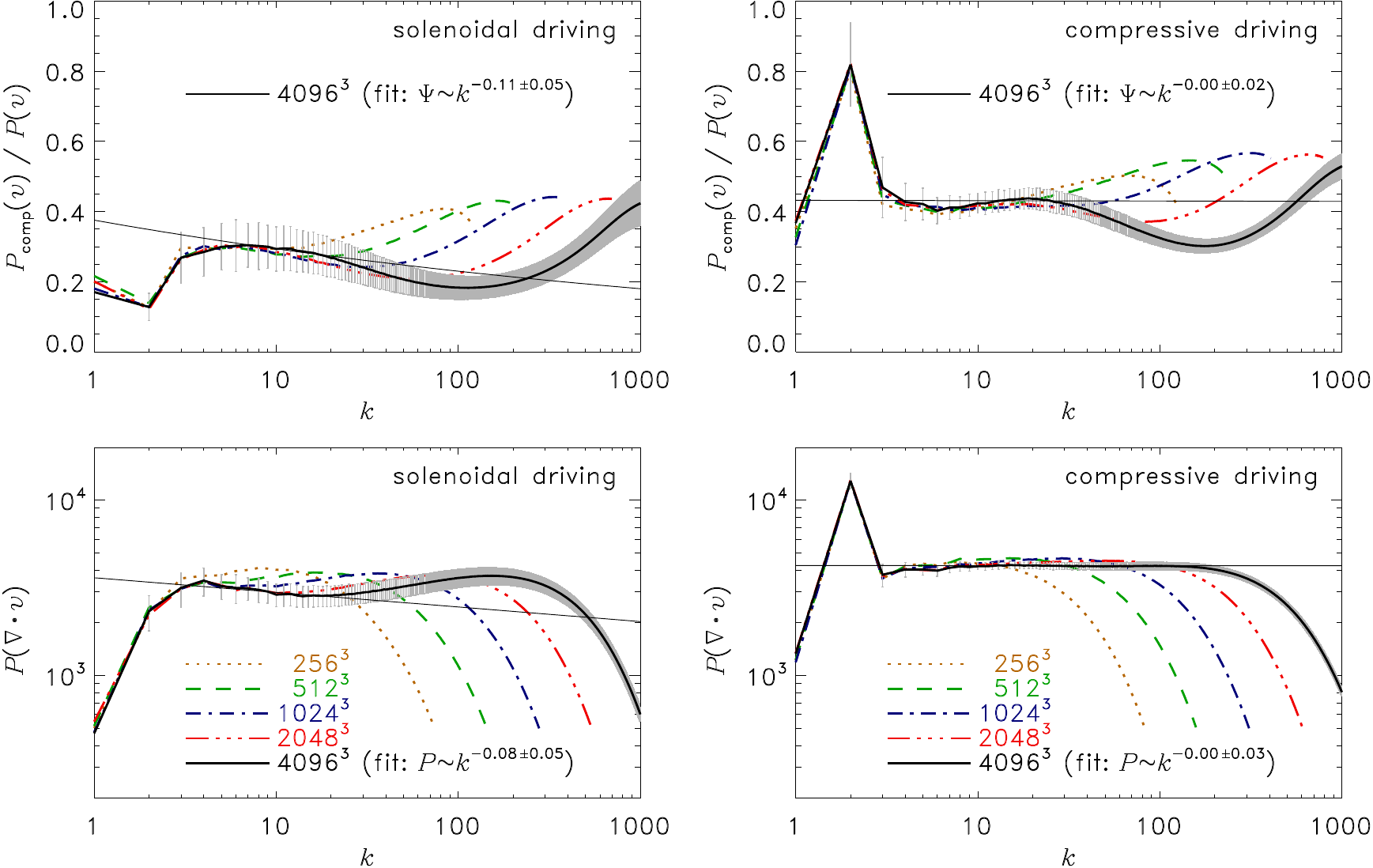}}
\caption{As Figure~\ref{fig:spect}, but showing Fourier power spectra of the compressive ratio defined in Equation~(\ref{eq:compratio}) (top panels) and of $\nabla\cdot\vect{v}$ (bottom panels) for solenoidal driving (left-hand column) and compressive driving (right-hand column). $\Psi$ and $P(\nabla\cdot\vect{v})$ decrease for solenoidal driving, but stay remarkably constant for compressive driving in the inertial range.}
\label{fig:spect_divv}
\end{figure*}

A detailed analysis of the scaling of the term $\mathcal{S}$ is beyond the scope of this study, but we can nevertheless quantify the amount of compression and $\nabla\cdot\vect{v}$, causing the modified scaling of $P(\rhothreev)$. To study the effects of compression, we first consider the compressive ratio spectrum,
\begin{equation} \label{eq:compratio}
\Psi(k)\equiv P_\mathrm{comp}(v)/P(v),
\end{equation}
i.e., the ratio of the longitudinal part of the velocity spectrum $P_\mathrm{comp}(v)$ (for which $\widehat{\vect{v}}(\boldsymbol{k})$ is parallel to $\boldsymbol{k}$) divided by the total velocity spectrum $P(v)$. The compressive ratio is a useful measure to evaluate the fraction of compressible velocity fluctuations as a function of scale. It is shown in Figure~\ref{fig:spect_divv} (top panels). We clearly see the effect of the distinct driving at $k=2$. Solenoidal driving does not excite compressible modes directly, producing a minimum in $\Psi(k)\approx0.1$ (note that this is not exactly zero, because some compression is indirectly induced at $k=2$, because the flow is supersonic). In contrast, compressive driving excites only compressible modes at $k=2$ and $\Psi(k)\approx0.8$ (it is also not exactly unity, because of indirect production of solenoidal modes in shock collisions and along density gradients). However, the \emph{direct} effect of the driving is only noticeable on scales $1 \leq k \leq 3$ (see Section~\ref{sec:methods}). Yet, we will see below that the driving does \emph{indirectly} change the statistics in the inertial range of compressible turbulence, as the supersonic turbulent fluctuations cascade down to smaller scales.

Figure~\ref{fig:spect_divv} (top panels) shows that the compressive ratio $\Psi$ decreases from about 1/3 to 1/4 for solenoidal driving in the inertial range, consistent with \citet[][]{KritsukEtAl2010}, but remains almost constant at $\Psi=0.43\pm0.04$ for compressive driving up to $k\approx30$, where the scaling is nearly converged with resolution (higher wave numbers, $k>30$, are affected by numerical dissipation and the bottleneck effect as explained above, and were thus excluded from the fit). This emphasizes the very different nature of supersonic turbulence driven by a solenoidal force and driven by a compressive force. In contrast to the classical concept of incompressible turbulence with an inertial range that does not depend on the driving, we see here that the inertial-range scaling of compressible turbulence depends on the driving.

Consistent with the compressive ratio spectrum, also $P(\nabla\cdot\vect{v})$ has a significant dependence on the driving as shown in the bottom panels of Figure~\ref{fig:spect_divv}. $P(\nabla\cdot\vect{v})\propto k^{-0.08}$ decreases in the inertial range for solenoidal driving. Remarkably though, for compressive driving, $P(\nabla\cdot\vect{v})$ remains constant over an extremely extended range of scales. We believe that this is the key reason for the different scaling of the $\rhothreev$ spectra. A quantitative analysis, however, requires a direct measurement of the scaling of $\varepsilon_\mathrm{eff}\propto\mathcal{S}(r)$, which must be done in a follow-up study.

\section{Conclusions} \label{sec:conclusions}
We studied the statistics of isothermal, highly compressible, supersonic (Mach 17) turbulence, such as relevant for the dynamics of the interstellar medium, with the Mach numbers of the order of \mbox{$5$--$20$} in the nearly isothermal density regime of molecular clouds. We analysed simulations with resolutions of \mbox{$256^3$--$4096^3$} grid cells to study the convergence of our results. Comparing the two limiting cases of turbulent driving: (1) by a solenoidal (divergence-free) force, and (2) by a compressive (curl-free) force, we find significant differences in the production of vorticity, the density PDF, and the scaling of the turbulence in the inertial range.

The vorticity produced by solenoidal driving is about $1.8\times$ larger than by compressive driving, close to the hypersonic limit (Figure~\ref{fig:tevol}). The turbulent structures, in particular the density structures, exhibit significantly different fractal dimensions for solenoidal and compressive driving (Figures~\ref{fig:images} and~\ref{fig:images_proj}). 
For \mbox{$1024^3$--$4096^3$} resolutions, the density PDFs are converged to within 20\% of the infinite-resolution limit, while simulations with $N_\mathrm{res}\leq256$ resolution can deviate by more than a factor of two from the infinite resolution limit (Figures~\ref{fig:pdfs} and~\ref{fig:pdfconv}). Compressive driving is more intermittent than solenoidal driving, with the PDF intermittency parameter $\theta_\mathrm{comp}/\theta_\mathrm{sol}\approx2.1$.

The pure velocity spectra are close to Burgers scaling with $P(v)\propto k^{-2}$ for both driving cases. In contrast, we find that the previously suggested universal scaling of the density-weighted velocity $\rhothreev$ is ruled out (see Figures~\ref{fig:spect} and~\ref{fig:spectevol}). The power spectrum $P(\rhothreev)\propto k^{-1.74\pm0.05}$, close to (but slightly steeper than) K41 scaling ($P\propto k^{-5/3}$) for solenoidal driving, is consistent with previous studies. However, $P(\rhothreev)$ is significantly steeper for compressive driving with $P(\rhothreev)\propto k^{-2.10\pm0.07}$ in the inertial range. The latter is in excellent agreement with the theoretical estimate $P\propto k^{-19/9}$ by \citet{GaltierBanerjee2011} for highly compressible turbulence with a strong $\nabla\cdot\vect{v}$ component, which we find to decrease for solenoidal driving, but stays almost perfectly constant for compressive driving down to very small scales (Figure~\ref{fig:spect_divv}).

Our study emphasizes the need to rethink the definition of the inertial-range scaling in highly compressible turbulence compared to incompressible turbulence. The latter defines an inertial range on scales sufficiently separated from the driving and dissipation scales, such that there is no influence of driving and dissipation. This basic rule cannot be carried over to highly compressible, supersonic turbulence, where the inertial-range scaling does depend on the driving, as we have shown here. This may be caused by supersonic turbulent fluctuations (shocks) crossing multiple scales, in contrast to the more local energy transfer between scales in incompressible turbulence. Answering this question requires a scale-by-scale analysis of the energy transfer tensor with high-resolution data in future studies.

\section*{Acknowledgements}
The author thanks Hussein Aluie, Supratik Banerjee, Sebastien Galtier, Ralf Klessen, Lukas Konstandin, Alexei Kritsuk, Paolo Padoan, Alessandro Romeo, and Rahul Shetty for stimulating discussions. An independent Bayesian analysis of the Fourier spectra by Lukas Konstandin is greatly appreciated. We thank the anonymous referee for a careful and critical reading, which significantly improved this study.
This work was supported by a Discovery Projects Fellowship from the Australian Research Council (grant DP110102191).
The simulations consumed computing time at the Leibniz Rechenzentrum (grant pr32lo) and the Forschungszentrum J\"ulich (grant hhd20).
The software used here was in part developed by the DOE-supported ASC/Alliance Center for Astrophysical Thermonuclear Flashes at the University of Chicago.

\appendix

\section{Velocity PDFs and time stepping} \label{app:vpdfs}

\begin{figure}
\centerline{\includegraphics[width=0.95\linewidth]{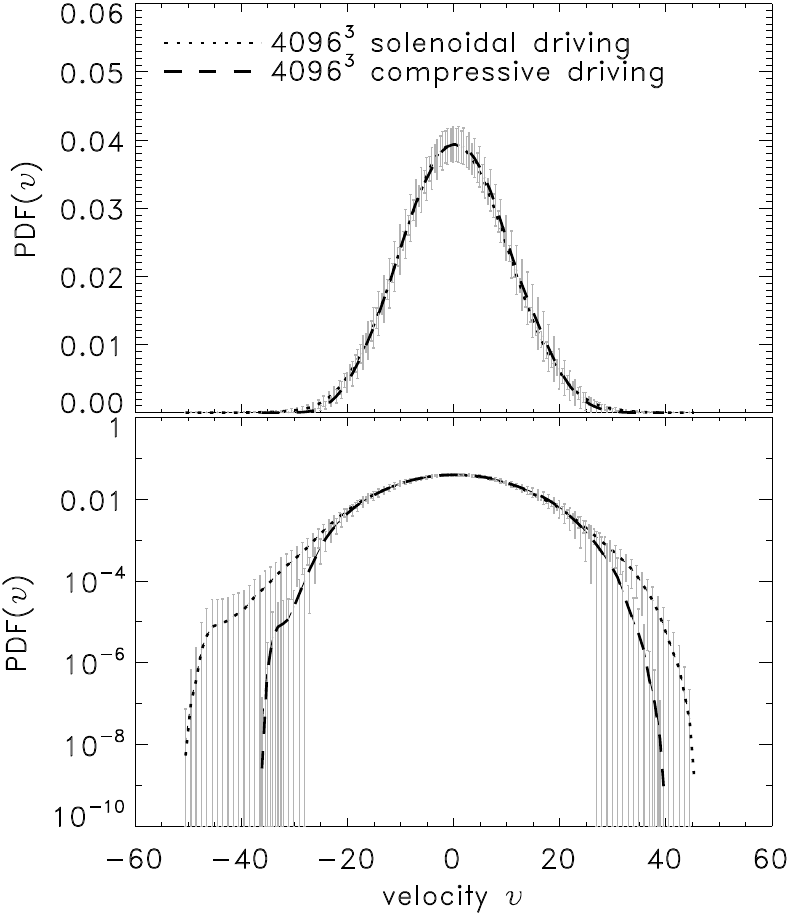}}
\caption{Time-averaged velocity PDFs (linear and logarithmic PDF axes in the top and bottom panels, respectively) for solenoidal driving (dotted line) and compressive driving (dashed line). The error bars show $1\sigma$ time variations and variations between the $x$, $y$, and $z$ components of the velocity. The $v$-PDFs are very close to Gaussian distributions and show maximum velocities (Mach numbers) of $v_\mathrm{max}\approx\pm50$.
}
\label{fig:vpdfs}
\end{figure}

Figure~\ref{fig:vpdfs} shows the PDFs of velocity $v$ averaged over all three spatial directions and averaged over time (error bars indicate $1\sigma$ temporal and spatial variations of the velocity components). Since we expressed the velocity in units of the sound speed throughout, maximum Mach numbers reach absolute values of about 50. The standard deviation of the $v$-PDF is practically identical to the rms Mach number, because the mean time-averaged velocity (and the mean momentum) is zero to machine precision.

Given these maximum velocities $|v_\mathrm{max}|\approx50$, typical time steps for the simulations with $N_\mathrm{res}=4096$ are $\Delta t \approx f_\mathrm{CFL}\,\Delta x / |v_\mathrm{max}| \approx 4\times10^{-6}$ for $\Delta x=1/4096$ and the CFL safety factor $f_\mathrm{CFL}=0.8$ \citep{CourantFriedrichsLewy1928}. The total number of time steps to evolve the simulations for six turbulent crossing times $6T$, where $T=L/(2\cs\mach)=1/(2\times17)\approx2.9\times10^{-2}$, is thus $N_\mathrm{steps} = 6T/\Delta t \approx 44,000$.

\end{document}